\documentstyle[editedvolume,numreferences,epsf]{crckapb} 

\newcommand{\keyword}[1]{\index{#1}#1}
\def\ket #1{|#1\rangle}
\def\bra #1{\langle #1 |}
\def\braket #1#2{\langle#1|#2\rangle}

\begin{opening}
\title{PHASE SEPARATION IN THE 2D HUBBARD MODEL\protect\\
       A CHALLENGING APPLICATION OF FIXED-NODE QMC}
       
\author{GIOVANNI B. BACHELET AND ANDREA C. COSENTINI}
\institute{Dipartimento di Fisica and INFM, Universit\`a La Sapienza,\\
           Piazzale Aldo Moro 2, I-00185 Roma, Italy}

\end{opening}

\runningtitle{PHASE SEPARATION IN THE 2D HUBBARD MODEL}
\makeindex
\begin{document}
\index{First Bachelet}

\section{Introduction}

A quantitative first-principles theory of atoms, molecules, and solids 
requires an accurate description of the electron-electron correlation.  
Fahy's \cite{fahy_book}, Mit{\'a}{\v s}' \cite{mitas_book}, and 
Umrigar's \cite{cyrus_book} contributions to this book suggest that 
Variational or Diffusion Monte Carlo simulations are rapidly 
approaching the required physical or chemical accuracy.  However, at 
least at their present pioneering stage, such simulations are still 
limited to relatively simple systems, whose correlation energy is a 
small fraction of the total energy, and, more significantly, whose 
main physics and chemistry can be qualitatively understood without any 
reference to many-body theory.  By contrast (and by definition), 
strongly correlated electron systems are those for which even a 
qualitative understanding of the main physics and chemistry is beyond 
the independent-electron picture; in this case, as also pointed out by 
Koch in his contribution to this book \cite{koch_book}, rather crude 
models of real interacting electrons, like the \keyword{Hubbard 
model}, can be on one hand the only feasible target, and, on the 
other, a clever way to capture the essential physics of a whole class 
of materials without going into too many realistic details 
\cite{realism}.  However, apart from very special limiting cases, even 
simple models of {\it interacting} electrons cannot be solved exactly, 
because, in spite of many severe approximations, they still give rise 
to an astronomically large \keyword{Hilbert space}; that's why quantum 
Monte Carlo methods, based on the idea of stochastic sampling, remain 
a natural option.  In this new context, however, the absence of a 
preliminary qualitative understanding of the electronic correlations 
makes the choice of the trial wavefunction (Sect.~\ref{FNMC}), and 
possibly even of the size of the simulation box (Sect.~\ref{Conclu}), 
a much more crucial and less straightforward part of the 
investigation.  A recent work by Hood {\it et al.} \cite{randy_hood} 
gives the occasion of an instructive glance at the state of the art, 
illustrating the different theoretical challenges involved in today's 
QMC studies of ``realistic'' and ``model'' systems.  Hood's 
variational Monte Carlo investigation of bulk silicon, based on 
Jastrow-Slater wavefunctions and a 54-atom 3D lattice, is more than 
adequate to obtain an essentially exact ground state, thereby 
explaining the great success of approximate local-density functionals 
for this kind of materials (similar considerations apply e.g.  to 
$s$-$p$ bonded materials, like boron nitride 
\cite{fahy_book,malatesta}).  By contrast, when investigating the 
Hubbard model, the extensive Fixed-Node Monte Carlo simulations by 
Cosentini {\it et al.} \cite{cosentini}, based on the nodal structure 
of a correlated Gutzwiller-Slater wavefunction and a (proportionally 
much larger) 256-atom 2D lattice, may in fact approach the true ground 
state of the infinite 2D Hubbard lattice with a much better accuracy 
than most previous attempts, and yet may not be sufficient to proclaim 
the ultimate word on \keyword{phase separation}.  Rather than giving a 
final answer to this really tough problem, the study presented in the 
second part of this chapter (Sects.~\ref{2DHub} and \ref{Conclu}) thus 
serves the more modest purpose of illustrating a particularly 
challenging application of the Fixed-Node quantum Monte Carlo method, 
recalled in the next Sect.~\ref{FNMC}.

\section{Fixed-Node QMC for lattice fermions}
\label{FNMC}

In what follows we briefly illustrate some aspects of the Fixed-Node 
QMC scheme actually followed in the study of the one-band Hubbard 
model of the subsequent Sects.~\ref{2DHub} and \ref{Conclu}.  We omit 
all demonstrations, which can be found either in other chapters of 
this book or in the original papers (in particular Ref.~\cite{CASO98} 
for the \keyword{power method} and Ref.~\cite{ceperley} for the 
\keyword{fixed-node} approximation).

\subsection{Power method on a lattice}
\label{powermethod}

Lattice models of quantum many-body systems (see e.g.  
Ref.~\cite{fisher}) are normally based on the preliminary choice of a 
tiny single-body basis set attached to each site of a lattice (for 
example, just one orbital per site in the simplest Hubbard model, or 
one spin per site in the Heisenberg model).  On a finite lattice of 
$N_{s}$ sites, the corresponding many-body Hilbert space is also 
discrete and finite, but its size is a very rapidly growing function 
of $N_{s}$ (for example, it's exponential for Heisenberg, 
combinatorial for Hubbard).  Since the purpose of lattice hamiltonians 
is to model infinite solids, small-$N_{s}$ lattices (which can be 
either analytically solved or exactly diagonalized on the computer) 
may be interesting as a benchmark for new theories and numerical 
methods (see e.g.  Table~\ref{table1}), but the real challenge is to 
reliably extrapolate the behavior of the infinite model solid from 
lattices of reasonably large $N_{s}$.  As mentioned, this amounts to 
dealing with finite but huge Hilbert spaces, where quantum Monte Carlo 
methods are often the only available alternative.

\subsubsection{A sparse matrix whose elements are all positive}
\label{negatoff}

In this context it is possible to project the ground state out of an 
arbitrary initial state (not orthogonal to it) by the ``power method'' 
-- the repeated application of some operator $\cal G$ simply related 
\cite{digression,excited} to the hamiltonian $\cal H$ (often 
conventionally called Green's function) -- provided that, in the 
(huge) many-body Hilbert space of the lattice model, one can choose a 
representation such that (a) each of the basis states $\ket{X}$ is 
connected only to a few other basis states $\ket{X'}$ by the operator 
$\cal G$ (or, in other words, most off-diagonal matrix elements of 
$\cal G$ are zero and the matrix $\bra{X'}{\cal G}\ket{X}$ is 
sparse), \index{sparse matrix} and (b) all the nonzero matrix elements of $\cal G$ 
are positive $\bra{X'}{\cal G}\ket{X} \geq 0$.  As described by 
Nightingale's contribution \cite{night_book}, the condition (a) is 
needed for the action of the hamiltonian $\cal H$ (or related 
\cite{digression} operators $\cal G$) on some arbitrary state to be 
computationally feasible -- the number of operations involved must be 
much smaller than the size of the Hilbert space \cite{sparse}; the 
condition (b) implies that, in the chosen representation, 
$\bra{X'}{\cal G}\ket{X}$ can be split into a \keyword{Markov matrix} 
and a positive weight, and stochastic approaches become possible (see 
e.g.  Ref.~\cite{night_book}); it also implies that the ground-state 
wavefunction \cite{excited}, in that particular representation, is 
positive everywhere $\braket{X}{\Psi_{\rm o}} \geq 0$, which rules out 
systems with three or more fermions ({\it sign problem}).  In other 
words, the power method presented here does not apply to the ground 
state of lattice fermions unless some {\it approximation} (recalled in 
Subs.~\ref{signproblem}) is adopted; it instead directly applies to 
those lattice models (like e.g.  the Heiseberg model on a square 
lattice \cite{CASO98}) for which both conditions (a,b) are satisfied.

\subsubsection{Markov chain (no weights)}
\label{markov}

Let us first suppose for purely pedagogical reasons that, besides (a) 
being sparse and (b) having all elements greater or equal to zero, the 
relevant matrix also (c) has the additional, remarkable property 
$\sum_{X'}\bra{X'}{\cal G}\ket{X}=1$.  If (c) holds, then 
$\bra{X'}{\cal G}\ket{X}$ can be directly identified with a 
probability (Markov) matrix $P_{X'X}$, whose maximum right eigenvector 
$\ket{\phi_{\rm max}}$ (i.e.  the right eigenvector corresponding to 
its largest eigenvalue $\lambda_{\rm max}=1$) can be first approached, 
and then sampled, by a (sufficiently long) single random walk, which 
can be generated on the computer by e.g.  the following simple rule:

\begin{itemize}
\item[1)] take an initial state $\ket{X_{\rm o}}$
\item[2)] find all the $N_{\rm o}$ states $\ket{X'}$
      for which $P_{X'X_{\rm o}}$=$\bra{X'}{\cal G}\ket{X_{\rm o}} \neq 0$
      (including $X'$=$X_{\rm o}$)
\item[3)] evaluate $P_{X'X_{\rm o}}=
      \bra{X'}{\cal G}\ket{X_{\rm o}} > 0$ for each of the 
	  above states, obtaining a list of 
	  $N_{\rm o}$ probabilities which add up to 1
\item[4)] align them as $N_{\rm o}$ adjacent segments adding up to form a 
	  unit segment
\item[5)] take a random number between $0$ and $1$ and, depending where 
      it falls on that unit segment, choose the corresponding state 
      $\ket{X'}$ as the new initial state $\ket{X_1}$
\item[6)] go back to the first item of this rule and iterate $K$ times 
	  this process (with very large $K$)
\end{itemize}

\noindent The above rule amounts to starting from an initial state 
$\ket{X_{\rm o}}$ and to repeatedly ($K$ times) multiplying from the 
left the matrix $P_{X'X}$.  The matrix-vector multiplication is 
implemented each time in a statistical sense: for example in the first 
step, rather than yielding a linear combination of all the $N_{\rm o}$ 
states $\ket{X'}$ with coefficients $P_{X'X_{\rm o}}$, the statistical 
result of the multiplication is just one of them, chosen at random 
among all the possible $N_{\rm o}$ states $\ket{X'}$ according to the 
probabilities $P_{X'X_{\rm o}}$.  After a good number $n$ of such 
stochastic matrix multiplications, the random sequence $ 
\ket{X_{n+1}}, \ket{X_{n+2}}, \ldots \ket{X_K}$ starts to be 
distributed according to $\ket{\phi_{\rm max}}$.  If the condition (c) 
holds, then the repeated multiplication of $P_{X'X}$ and the repeated 
application of the operator $\cal G$ are the same thing, 
$\ket{\phi_{\rm max}}$ is proportional to $\ket{\Psi_{\rm o}}$, and 
the random walk $ \ket{X_{n+1}}, \ket{X_{n+2}}, \ldots \ket{X_K}$ can 
be directly used for statistical averages over the ground state.

\subsubsection{Single random walk (weights)}

Unfortunately general Markov matrices $P_{X'X}$ are not symmetric 
(that's why one has to distinguish between left and right 
eigenvectors), and $\cal H$ or related hermitian operators $\cal G$ do 
not satisfy the condition (c) of previous Subs.~\ref{markov}: one has, 
instead, $\sum_{X'}\bra{X'}{\cal G}\ket{X}=b_{X} \neq 1$.  However, if 
the property (b) does hold, then $b_{X} > 0$, and we can still 
decompose $\bra{X'}{\cal G}\ket{X}$ into the product of a normalized 
probability matrix $P_{X'X}$ and a positive weight $b_{X}$.  Then, by 
the simple rule illustrated above (steps 1-6) we can still generate a 
random walk which samples the maximum right eigenvector of 
$P_{X'X}=\bra{X'}{\cal G}\ket{X}/b_{X}$, except that now the 
eigenvector $\ket{\phi_{\rm max}}$ is no longer proportional to 
$\ket{\Psi_{\rm o}}$, the ground-state eigenvector of $\cal H$.  
However, if we keep record not only of the state $\ket{X_j}$, but also 
of the weight $b_{j}$ = $b_{X_{j}}$ = $\sum_{X'}\bra{X'}{\cal 
G}\ket{X_{j}}$ (which must be calculated at each step because it's 
needed to obtain a Markov matrix $P$ from the original matrix $\cal 
G$), then the random walk $\{ \ket{X_i} \}$ ($n$$<$$i$$\leq$$K$) which 
samples $\ket{\phi_{\rm max}}$ can also be used to sample ${\cal 
G}^{L}\ket{\phi_{\rm max}}$: one only needs to associate an $L$-th 
order cumulative weight $w_i(L)=b_{i-L}b_{i-L+1}b_{i-L+2}\ldots 
b_{i-1}$ to each state $\ket{X_{i}}$ of the sequence, and then 
consistently include these weights in the averages \cite{CASO98}.  
Since the repeated application of $\cal G$ tends to project the 
ground-state component out of $\ket{\phi_{\rm max}}$, a sufficiently 
large $L$ could allow ground-state averages from a single random walk 
\cite{averages}.  Unfortunately this is not possible because, when $L$ 
is large enough for ${\cal G}^{L}$ to project the ground state out of 
$\ket{\phi_{\rm max}}$, then the weights $w_i(L)$ -- as suggested by 
their product form and argued in better detail elsewhere \cite{CASO98} 
-- undergo wild fluctuations, which grow exponentially with $L$.  So 
statistical averages are accompanied by a diverging variance and 
become pointless.

\subsubsection{Statistical averages, many walks, reconfiguration}

The standard solution to the problem of a diverging variance, which 
occurs with a single random walk, is to propagate many simultaneous 
random walks.  These walks propagate independently most of the time, 
but not all the time (else they would be equivalent to a single, long 
walk): to prevent individual weights from getting too small or too 
large and the variance from blowing up, the different walks must 
periodically undergo some sort of global reshoveling of states and 
weights.  As recently emphasized by Calandra and Sorella 
\cite{CASO98}, it's this {\it reconfiguration} process which makes the 
difference.  The results of Sects.~\ref{2DHub} and \ref{Conclu} are 
actually based on their new \keyword{reconfiguration} scheme.  A 
thorough discussion of the underlying theory, and even a complete 
presentation of their recipe, is beyond the scope of this chapter; it 
can be found by the interested reader in their recent, original paper 
\cite{CASO98}.  But our numerical experience may be worth mentioning: 
in practice the Calandra-Sorella reconfiguration scheme allowed us to 
reproduce with a relatively small fixed number ($100 \sim 200$) of 
walkers equally accurate results as those obtained by means of 
standard branching schemes \cite{trivedi} with more than $2000$ 
walkers.

\subsection{The sign problem}
\label{signproblem}

As mentioned in the previous Subs.~\ref{powermethod}, the power method 
can work as a ground-state \cite{excited} projector only if all the 
matrix elements of the operator $\cal G$ are positive 
\cite{digression}, or, equivalently, if the off-diagonal matrix 
element of the hamiltonian are all negative
\cite{digression,excited}.

\subsubsection{Fermion nodes and other possible sources}
\label{sources}

The above condition on matrix elements can be violated for various 
reasons: (i) particle statistics: the ground-state wavefunction of 
identical fermions changes sign under their exchange, and must be 
negative somewhere; (ii) one-particle effects: in certain lattice 
models, some hopping terms have the ``wrong'' sign (on continuum 
``realistic'' models a similar problem may arise from nonlocal 
\keyword{pseudopotentials} \cite{hsc} as illustrated by Mit{\'a}{\v s} 
in this book \cite{mitas_book}); (iii) other effects, like the 
frustration of a quantum spin system on a triangular lattice 
\cite{boninsegni}; (iv) combinations of the above.  From this point of 
view our simplest 2D one-band Hubbard model on a square lattice falls 
in the first category (i), since its \keyword{sign problem} is 
entirely due to the Fermi statistics of the electrons 
\cite{secondqua}.

\subsubsection{Importance sampling and Fixed-Node approximation}
\label{fissanodi}

As clearly explained in at least one of the other chapters of this 
book \cite{night_book}, the use of importance sampling is of crucial 
importance to improve the sampling efficiency; a trial function may 
also be the simplest way to incorporate the Fermi statistics in a 
basis of ``labeled fermion'' lattice configurations 
\cite{ceperley,secondqua}.  So in actual fermion calculations a good 
trial function $\Psi_{T}$ (typically available from preliminary 
Variational Monte Carlo calculations) is always adopted.  In the power 
method, importance sampling amounts to replacing $\cal G$ by 
$\tilde{\cal G}$, whose matrix elements are $\tilde{\cal 
G}_{X'X}=\Psi_{T}(X') \bra{X'}{\cal G} \ket{X}\Psi_{T}^{-1}(X)$, and 
then to consistently deal with $\tilde{\cal G}$ rather than $\cal G$ 
\cite{night_book}.  The sign problem amounts then to the occurrence of 
some unwanted $\tilde{\cal G}_{X'X} < 0$, and, apart from transient 
estimates which are rather hazardous for large systems, the only cure 
to this problem seems to be the fixed-node approximation recently 
proposed by An, Bemmel, Ceperley, Haaf, Leeuwen, and Saarloos 
\cite{ceperley,bemmel}.  In this approximation the true hamiltonian 
$\cal H$ is replaced by an effective hamiltonian ${\cal H}_{\rm eff}$, 
defined as follows:

\begin{eqnarray}
\bra{X'} {\cal H}_{\rm eff} \ket{X} & = &
0 \hspace*{7.8em} \mbox{if $\tilde{\cal G}_{X'X} < 0$} \nonumber
\\ & = &
\bra{X'} {\cal H} \ket{X}
\hspace*{4.1em} \mbox{(otherwise)}
\end{eqnarray}

\noindent for the off-diagonal terms $X \neq X'$, so that hops from 
the state $\ket{X}$ towards ``sign-flipping states'' $\ket{X'}$ (i.e.  
such that $\tilde{\cal G} < 0$) are now forbidden, and, for the 
diagonal terms,

\begin{equation}
\bra{X} {\cal H}_{\rm eff} \ket{X} = \bra{X} {\cal H} \ket{X} +
\sum_{X'}^{sf}
\bra{X'} {\cal H} \ket{X}\frac{\Psi_{T}(X')}{\Psi_{T}(X)},
\end{equation}

\noindent the summation being over all the ``sign-flipping states'' 
$\ket{X'}$ such that $\bra{X'}{\cal H}\ket{X} \neq 0$ but $\tilde{\cal 
G}_{X'X} < 0$: whenever in the ``neighborhood'' of the state $\ket{X}$ 
there are sign-flipping states $\ket{X'}$, appropriate repulsive terms 
are added to the old diagonal element $\bra{X}{\cal H}\ket{X}$.  Such 
a diagonal repulsive term has the intuitive effect of repelling the 
random walk away from the regions of the Hilbert space where 
$\tilde{\cal G}$ would tend to revert its sign, and also the less 
intuitive, but more important effect of producing, in analogy with the 
continuum case \cite{fixednode,upbound}, an upper bound for the true 
ground-state energy and a variational approximation \cite{ceperley}.  
This formulation, contained in Ref.\cite{ceperley}, extends the 
original prescription of Ref.~\cite{bemmel} to include all the 
possible sources of a sign-flip in $\cal G$ (Subs.~\ref{sources}) 
\cite{lever}.  However for our simple Hubbard hamiltonian 
(Eq.~\ref{hubbardham}) the original and new prescription are 
coincident, because in this case the only source of sign flips are the 
Fermi nodes \cite{secondqua}.  Although the original papers by An, 
Bemmel, Ceperley, Haaf, Leeuwen, and Saarloos \cite{ceperley,bemmel} 
are only 3-4 years old, their ``fixed-node Monte Carlo for lattice 
fermions'' has been already used by other authors, like Boninsegni for 
frustrated Heisenberg systems \cite{boninsegni}, or Koch {\it et al.} 
for orbitally-degenerate Hubbard models \cite{koch_book,gunnarsson}.  
We (Cosentini {\it et al.} \cite{cosentini}) have chosen it for the 
extensive study of the 2D Hubbard model presented in the coming 
Sections \ref{2DHub},\ref{Conclu}, because it allows a reliable 
investigation of (previously unfeasible) very large lattice-fermion 
systems.

\section{2D Hubbard model and phase separation}
\label{2DHub}

\subsection{Motivations}
\label{motivi}

\subsubsection{High-$T_c$, phase separation, ICDW}

Strongly correlated electrons and holes are expected to play a key 
role in the \keyword{high-$T_c$ superconductors}.  Their possible 
instability towards phase separation, initially believed to inhibit 
superconductivity, is attracting a lot of interest since a few 
different authors \cite{clc,emery,dagotto} have pointed out that such 
a tendency may in fact be intimately related to the high-$T_c$ 
superconductivity.  Long-range repulsive interactions may turn the 
phase-separation instability into an incommensurate 
charge-density-wave (\keyword{ICDW}) instability, and the very 
existence of a quantum critical point associated with it may be a 
crucial ingredient of the superconducting transition \cite{clc2}.  
Phase-separation and/or ICDW instabilities are related to a 
substantial reduction of the kinetic energy, which otherwise tends to 
stabilize uniformly distributed states; such a reduction is typical of 
strongly correlated electrons, both in real and model systems.

\subsubsection{What is phase separation?}

Phase separation is a \keyword{thermodynamic instability} associated 
with the violation of the stability condition $\chi^{-1} = {\partial}^2 
\!{\cal E}/ \partial{n^2} > 0$, which requires the energy density 
${\cal E}$ of an infinite electronic system to be a convex function of 
the electron density $n$.  If such a violation occurs in the density 
range $n_1 < n < n_2$, then the system will separate into two 
subsystems with electron densities $n_1$ and $n_2$.  For the 
two-dimensional $t\!-\!J$ and Hubbard models a phase separation, if 
any, is expected to occur in a density range close to half filling ($n 
\simeq 1$), and to yield a hole-rich phase with electron density 
$n_1<1$ and a hole-free phase with electron density $n_2=1$ 
\cite{emery2}.  In view of our further developments, the same physics 
can be conveniently rephrased in terms of the hole density $x=1-n$ and 
energy density $e_{h}(x)={\cal E}(n(x))$, as done in the upper right 
panel of Fig.~\ref{fig0}.

\subsubsection{Maxwell's construction}

\begin{figure}
\begin{center}
\epsfxsize=200 pt
\centerline{\epsfbox{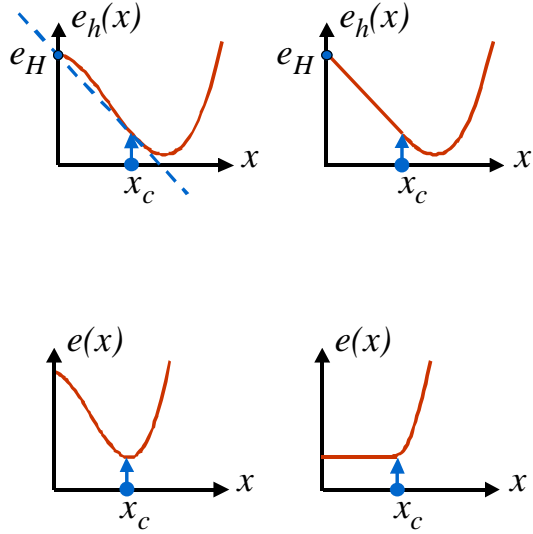}}
\end{center}
\caption{Sketch of a hypothetic system of holes which, when its 
density falls between zero and a critical value $x_c$, tends to 
phase-separate between a hole-free phase (zero density) and another 
hole-rich phase (density equal to $x_c$).  Upper panel: behavior of 
the hole energy density $e_h$ vs.  the hole density $x$ in the case of 
a finite (left) and an infinite (right) system.  For the finite system 
(left), between $x=0$ and $x=x_c$, the actual energy of the 
phase-separated system (dashed straight line, Maxwell's construction) 
is lower than the energy of the homogeneous system; for the infinite 
system, in the same density range, the energy density is a straight 
line, and the inverse compressibility vanishes.  Lower panel: behavior 
of the function $e(x)\!=\![e_h(x)\!-\!e_H]/x$ for a finite (left) and 
an infinite (right) system.  As shown by Emery {\it et al.}, in a 
finite system the study of $e(x)$ is completely equivalent to 
Maxwell's construction (see text).}
\label{fig0}
\end{figure}

In a truly infinite system a phase separation of the type just 
discussed would be associated to a vanishing inverse compressiblity 
$\chi^{-1}$ in the whole density range $n_1 < n < n_2$; in a finite 
system $\chi^{-1}$ may even become negative, because of surface 
effects; the same applies of course to the the second derivative of 
the hole energy density $e_{h}(x)$ with respect to the hole density 
$x$.  So for finite systems it's preferable to pinpoint the phase 
separation using \keyword{Maxwell's construction} (Fig.~\ref{fig0}, 
upper left panel).  It has been shown in Ref.\cite{emery2} that 
Maxwell's construction is equivalent to study, as a function of the 
hole density $x \!=\!1\!-\!n$, the quantity $e(x) \!=\!  
[e_{h}(x)\!-\!e_{H}]/x$, i.e.  the energy per hole $e_{h}(x)$ measured 
relative to its value at half filling $e_{H}=e_{h}(x \!=\!0)$ 
(Fig.~\ref{fig0}, lower panels).  For an infinite system, if the 
inverse compressibility $\chi^{-1}$ vanishes between some critical 
density $x_{c}$ and half filling $x=0$ (upper right), then the 
function $e(x)$ is a constant for $0 \leq x \leq x_{c}$, and the 
fingerprint of a phase separation is thus a horizontal plot of $e(x)$ 
below $x_{c}$ (lower right).  For a finite system, instead, the 
inverse compressibility $\chi^{-1}$ may become negative (upper left) 
and the fingerprint of a phase separation is a minimum of $e(x)$ at 
$x=x_{c}$ \cite{emery2} (lower left).  On finite lattices, where only 
a discrete set of densities is available and a statistical error bar 
usually accompanies the estimated QMC energies, the plot of the 
function $e(x)$ (lower panels of Fig.~\ref{fig0}) is supposed to allow 
an easy visual judgement of the presence or absence of such an 
instability.  The attempt of telling whether near half filling a few 
electron or hole energies are better interpolated by a straight or by 
a somewhat curved line (upper panels of Fig.~\ref{fig0}) is definitely 
more difficult than looking at $e(x)$.  However even $e(x)$ can give a 
reliable criterion only for medium-large finite systems; really small 
systems (for which most numerical results have been up to now 
available) can attain so few and coarse densities, and suffer from so 
large finite-size errors, that their predictions of the relevant 
trends remain largely inconclusive, no matter which energy function we 
look at.  More remarks on $e(x)$ vs.  phase separation will be 
presented later on, when examining the numerical results.

\subsubsection{Limited experimental evidence}
\label{experiment}

Phase separation has been experimentally observed in 
La$_2$CuO$_{4+\delta} $\cite{jorgensen88,chou96}, where the oxygen 
ions can move: in the doping interval $0.01\leq \delta\leq0.06$ the 
compound separates into a nearly stoichiometric antiferromagnetic 
phase and a superconducting oxygen-rich phase.  In generic compounds, 
where charged ions cannot move, the possibility of a macroscopic phase 
separation is spoiled by the long-range Coulomb repulsion, and should 
lead to an ICDW instability \cite{tranquada}; here, 
however, the identification of charge inhomogeneities with spoiled 
phase separation is less obvious \cite{bianconi}.

\subsubsection{Theoretical controversy}

On the theoretical side, evidence for phase separation has been 
suggested for various models of strongly correlated electrons, such as the 
\keyword{$t\!-\!J$ model} \cite{emery2}, the three-band Hubbard model, the 
Hubbard-Holstein model and the Kondo model (see e.g.  Ref.  
\cite{clc2} and references therein).  Despite intensive studies, even 
for simple models there is no general agreement on the phase 
separation boundary: for the very popular $t\!-\!J$ model, phase 
separation is fully established only at large $J$, but at small $J$ 
(which unfortunately happens to be the physically relevant case) 
theoretical and numerical results are quite controversial.  Emery {\it 
et al.}'s \cite{emery2} theory that phase separation occurs at {\it 
any} value of $J$ in the $t\!-\!J$ model is confirmed by a recent 
numerical study by Hellberg and Manousakis \cite{hellberg}, but is in 
contrast with Dagotto {\it et al.}'s \cite{dagotto} exact numerical 
results on small clusters, suggesting no tendency toward phase 
separation for both the Hubbard model and the $t\!-\!J$ model below a 
critical value $J < J_c \sim t$, and with Shih {\it et al.}'s 
\cite{taipei} numerical results.  We also mention the recent 
suggestion by Gang Su \cite{gangsu}, according to which the Hubbard 
model does not show phase separation for any value of $U/t$ at any 
temperature.

\subsubsection{A challenging application of fixed-node quantum Monte 
Carlo}

Given the above theoretical and experimental situation, the 
availability of the fixed-node quantum Monte Carlo, described in the 
above Subs.~\ref{powermethod}, has recently provided us (Cosentini 
{\it et al.} \cite{cosentini}) with a strong motivation to further 
investigate the Hubbard model.  Whether the 2D Hubbard hamiltonian, a 
prototype for interacting electrons with no long-range repulsion, 
shows any instability towards phase separation, is by itself an 
interesting open question.  Besides the intrinsic interest in the 
model's behavior, a numerical investigation of this question may also 
shed some indirect light on two related issues: (i) whether phase 
separation is likely to occur in the $t\!-\!J$ model at small $J$ (the 
``physical region'', where the ground-state energy difference with the 
Hubbard model should become negligible); and (ii) whether the measured 
phase separation of some real high-$T_c$ superconductors 
(Subs.~\ref{experiment}) is such a fundamental and elementary fact as 
to be explained by the simplest one-band Hubbard hamiltonian.

\subsection{Numerical strategy and tests}
\subsubsection{Hamiltonian}

To obtain numerical estimates of functions like those sketched in 
Fig.~\ref{fig0} we need to to evaluate the ground-state energy of the 
Hubbard hamiltonian
 \begin{equation}
\label{hubbardham}
{\cal H} = -t\sum_{\langle i,j 
\rangle\sigma}(c^{\dagger}_{i\sigma}c_{j\sigma} + h.c.)  + U\sum_{i} 
n_{i\uparrow}n_{i\downarrow}
\end{equation}
for many large finite lattices and many different electron densities.  
As already mentioned (and easily double-checked by the reader), for 
fermions this hamiltonian ${\cal H}$ and the corresponding 
importance-sampled $\tilde{\cal G}$ satisfies the requirement of 
sparseness (Subs.~\ref{negatoff} and Ref.~\cite{sparse}) but not the 
one on sign (Subs.~\ref{signproblem} and Ref.~\cite{secondqua}).  The 
choice of a good variational wavefunction is thus needed not only to 
improve the sampling efficiency, but also for the power method to 
become feasible through the fixed-node approximation 
(Subs.~\ref{fissanodi}).

\subsubsection{Choice of variational wavefunction}

\begin{figure}
\begin{center}
\epsfxsize=200 pt
\centerline{\epsfbox{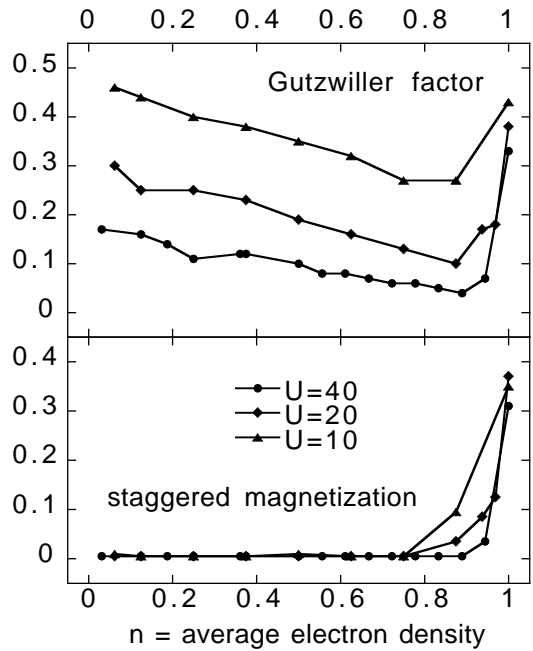}}
\end{center}
\caption{A collection of optimal Gutzwiller factors $0<g<1$ (upper 
panel) and staggered magnetizations $0<m<1$ (lower panel) for many 
different electron densities and three values of the coupling strength 
$U/t$: $10$ (triangles), $20$ (diamonds), and $40$ (dots).  Most 
results correspond to $16\!\times \!16$ lattices, but a subset 
corresponds to $12\!\times \!12$ and to other large but different 
lattices.  When the size is large enough, the optimal $g,m$ appear thus 
to be almost independent of the size.}

\label{stag-gutz}
\end{figure}

The variational wavefunctions we use to guide the random walks and to 
fix the nodes are the product of a Gutzwiller factor \index{Gutzwiller wavefunction} and two 
\keyword{Slater determinant}s of single-particle, mean-field 
wavefunctions (with uniform electron density $n$ and uniform staggered 
magnetization density $m$) for an equal number of up- and down-spin 
electrons \cite{singlet}.  Preliminary estimates of the optimal Gutzwiller
parameter $0<g<1$ 
and mean-field wavefunctions (parametrized by the staggered 
magnetization $0<m<1$) were obtained, for each $U/t$ and 
density, by variational Monte Carlo (VMC) runs; they are shown in 
Fig.~\ref{stag-gutz} as a function of the electron density $n$ for 
three choices of the coupling strength $U$ (in units of the hopping 
parameter $t$).  As $U/t$ increases, the optimal $g$ decreases, 
reflecting a stronger depression of doubly-occupied configurations.  
As far as the density dependence is concerned, the staggered 
magnetization $m$ is close to zero for all $U/t$ and low-medium 
electron densities, but sharply raises above $n\sim 0.75$, thus 
signaling, at the variational level, the system's tendency towards an 
antiferromagnetic ordering as half filling approaches.  The optimal 
Gutzwiller factor steadily decreases as a function of the electron 
density $n$, except for very high electron densities ($n\simeq 0.9$ 
and above), where it apparently goes up again, as also noticed 
elsewhere \cite{koch_priv}.  In this narrow density range, however, 
the energy minimum appears to be flat: slightly different $g$ values 
give essentially the same VMC energy, and the corresponding trial 
wavefunctions also yield the same energy at the fixed-node level.

\subsubsection{Comparison with exact results}

A few representative variational (VMC) and fixed-node Monte Carlo 
(FNMC) energies are shown in Table~\ref{table1} for the $4 \!\times\!  
4$ Hubbbard lattice, for which exact results \cite{sorella} are 
available.  As expected, the VMC energy is always above the fixed-node 
energy, which in turn, for these coupling strengths, is slightly 
($\sim 3\%$) above the exact energy.  For comparison we show the 
Constrained-Path Monte Carlo (CPMC) energies of Zhang {\it et al.} 
\cite{gubernatis,zhang}, which also include a larger $16 \!  \times \!  
16$ lattice (last row).  Especially at large $U/t$ our results 
appear to be of comparable quality to theirs.  As far as the $4 \!\times\!  
4$ results are concerned, we notice that for $N_{e}=10$, which 
corresponds to a closed-shell configuration, both fixed-node and CPMC 
are much closer to the exact energy than for $N_{e}=14$, which 
corresponds to an open-shell configuration.  This could be a serious 
problem when numerically studying the behavior of the energy as a 
function of the density; the results presented here suggest that, for 
lattices larger than $12\!\times\!12$, the shell effects become almost 
irrelevant \cite{maddeche}.

\begin{table}

\begin{center}
\begin{tabular}{cccccccccc}
\hline
& ${\rm size}$ & $N_{e}$ &
 $n$ &
 $U/t$ &
 ${\rm VMC}$  &
 ${\rm FNMC}$ &
 ${\rm CPMC}$ &
 ${\rm EXACT}$ &
\\
\hline
& $ 4 \!\times\! 4 $ & $ 10 $ &
$ 0.625 $ & $ 4 $ & $ -1.211(2) $ & $ -1.220(2) $ & $ -1.2238(6) $ & $
-1.2238 $ &\\
& $ 4 \!\times\! 4 $ & $ 10 $ &
$ 0.625 $ & $ 8 $ & $ -1.066(2) $ & $ -1.086(2) $ & $ -1.0925(7) $ & $
-1.0944 $ &\\
& $ 4 \!\times\! 4 $ & $ 14 $ &
$ 0.875 $ & $ 8 $ & $ -0.681(2) $ & $ -0.720(2) $ & $ -0.728(3) $ & $
-0.742$ &\\
& $ 4 \!\times\! 4 $ & $ 14 $ &
$ 0.875 $ & $12 $ & $ -0.546(2) $ & $ -0.603(2) $ & $ -0.606(5) $ & $
-0.628$ &\\
&
$ 16 \!\times\! 16 $ & $ 202 $ &
$ 0.789 $ & $4 $ & $ -1.096(2) $ & $ -1.107(5) $ & $ -1.1193(3) $ & $
- $ &\\
\end{tabular}
\end{center}
\vspace{-0.2 truecm} 
\caption{Ground-state energy per site (in units of the hopping 
parameter $t$) for a $4 \!  \times \!  4$ Hubbard lattice and various 
values of $U/t$.  $N_{e}$ is the number of electrons and $n$ is the 
corresponding average density.  VMC: variational Monte Carlo, 
Ref.~\protect\cite{cosentini}; FNMC: Fixed-Node Green's function Monte 
Carlo, Ref.~\protect\cite{cosentini}; CPMC: Constrained-Path Monte 
Carlo, Ref.~\protect\cite{gubernatis}; EXACT: exact diagonalization 
results, Ref.~\protect\cite{sorella}.(see text).}
\label{table1}
\end{table}

\subsubsection{How to vary the density?}

To study the energy as a function of the electronic density we have 
first tried out the less usual way of varying the density suggested in 
Ref.~\cite{hellberg} to ``avoid spurious Fermi-surface shape 
effects'': keep the number of electrons $N_{e}$ fixed while the size 
of the underlying lattice is varied.  But we discovered that by 
this prescription, if the number of electrons is really small (e.g.  
$N_{e}\!=\!16$), then artificial changes in the convexity of the curve 
may occur.  If, instead, the system is large enough (e.g.  
$12\!\times\!12$ lattices or larger), then it doesn't matter how the 
density is varied, as this prescription and the usual 
prescription give the same results.  So for our systematic study (many 
densities and three $U/t$ values) we finally adopted a large $16 
\!\times\!  16$ lattice ($N_{s}\!=\!256$ sites), and varied the number 
of electrons $N_{e}$ to yield electronic densities $n\!=\!N_{e}/N_{s}$ 
ranging from empty $n\!=\!0$ to half filling $n\!=\!1$.

\subsection{Results}
\label{linummeri}

\subsubsection{Energy vs. density for large lattices}

\begin{figure}
\begin{center}
\epsfxsize=200 pt
\centerline{\epsfbox{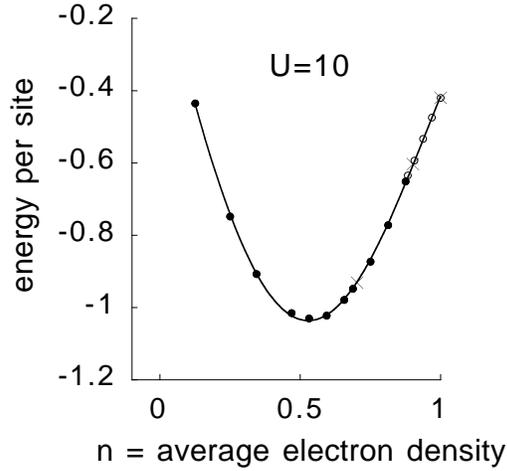}}
\end{center}
\caption[xxx]{Ground-state energy per site (in units of the hopping 
parameter $t$) as a function of the electronic density, for a 2D 
Hubbard lattice of $N_{s} = 16 \!\times\!  16 = 256$ sites with 
$U/t=10$.  Errors are smaller than the dot size.  Full dots correspond 
to closed shells and empty dots correspond to open shells.  The three 
crosses correspond to closed-shell densities of a $11\sqrt 2\times 
11\sqrt 2$ lattice, and are shown for comparison (see text and 
Refs.\protect\cite{apbc,shellstorte}).}
\label{fig1}
\end{figure}

\begin{figure}
\begin{center}
\epsfxsize=200 pt
\centerline{\epsfbox{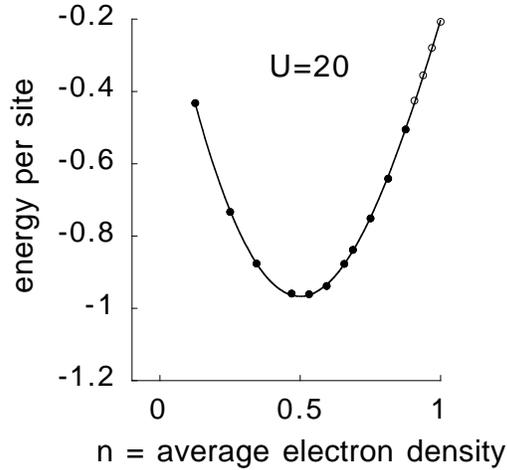}}
\end{center}
\caption{Ground-state energy per site (in units 
of the hopping parameter $t$) as a function of the electronic density, 
for a 2D Hubbard lattice of $N_{s} = 16 \!\times\!  16 = 256$ sites 
with $U/t=20$.  Errors are smaller than the dot size.  Full dots 
correspond to closed shells and empty dots correspond to open shells.}
\label{fig2}
\end{figure}

\begin{figure}
\begin{center}
\epsfxsize=200 pt
\centerline{\epsfbox{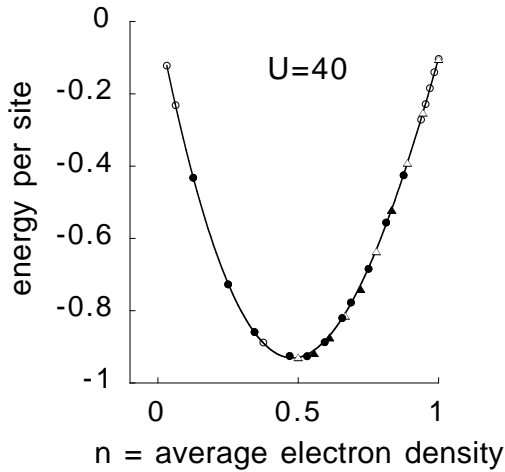}}
\end{center}
\caption{Ground-state energy per site (in units 
of the hopping parameter $t$) as a function of the electronic density, 
for a 2D Hubbard lattice of $N_{s} = 16 \!\times\!  16 = 256$ sites 
with $U/t=40$.  Errors are smaller than the dot size.  Full dots 
correspond to closed shells and empty dots correspond to open shells.  
Full and empty triangles, corresponding to closed and open shells of a 
smaller $12\!\times\!12$ lattice, are shown for comparison, and give 
a measure of the size effects (see text and 
Ref.~\protect\cite{maddeche}).}
\label{fig3}
\end{figure}

\begin{figure}
\begin{center}
\epsfxsize=200 pt
\centerline{\epsfbox{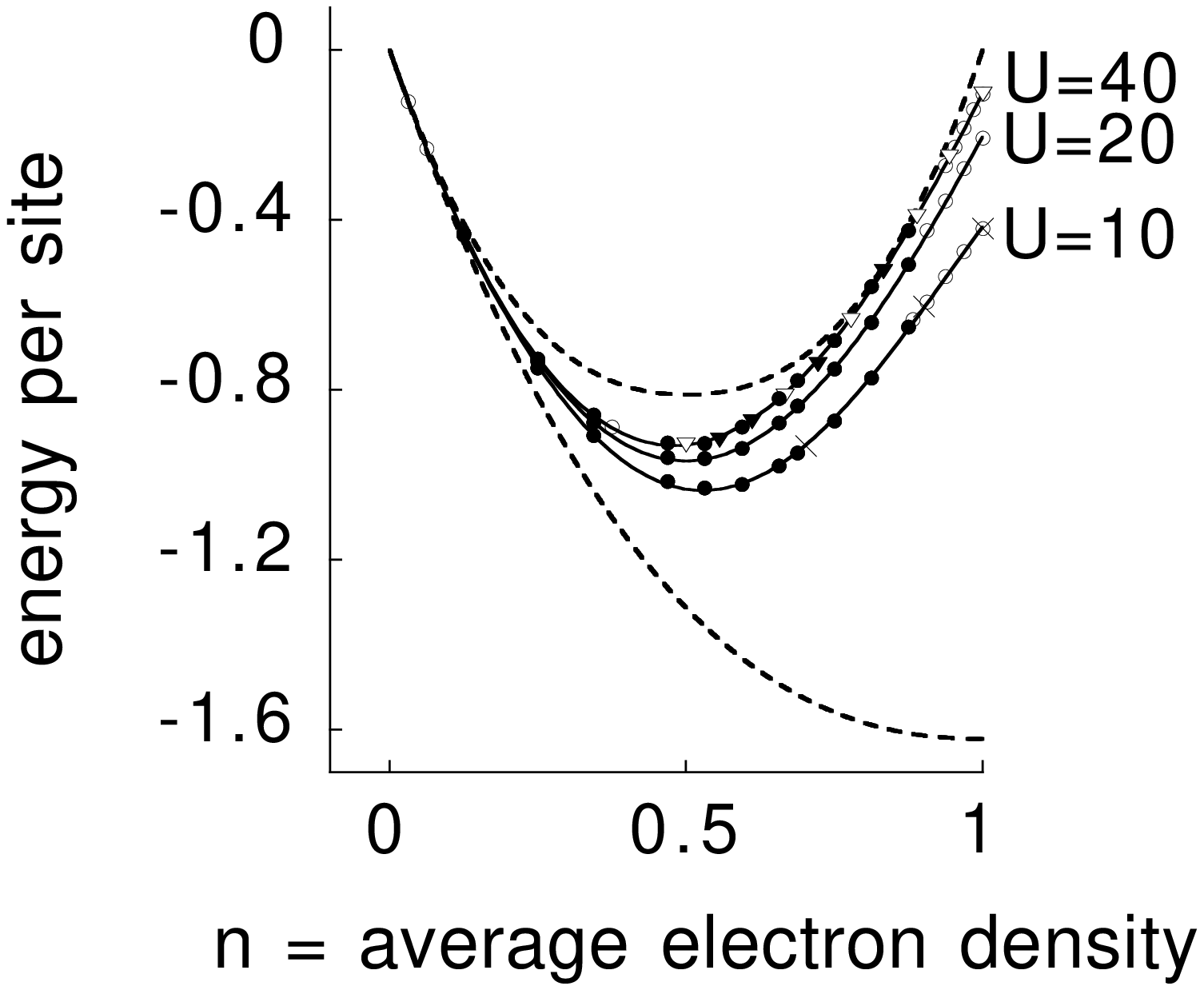}}
\end{center}
\caption{This picture summarizes the results of the previous 
Figs.~\protect\ref{fig1}-\protect\ref{fig3}, showing the ground-state 
energy per site (in units of the hopping parameter $t$) as a function 
of the electronic density, for a 2D Hubbard lattice of $N_{s} = 16 
\!\times\!  16 = 256$ sites with $U/t=10$ (lower), $20$ (middle), and 
$40$ (upper data).  Errors are smaller than the dot size and symbols 
are the same as in Figs.~\protect\ref{fig1}-\protect\ref{fig3}.  The 
dashed curves correspond to two known results for the infinite lattice 
at $U=0$: the fully spin-polarized case (upper curve), whose total 
energy per site is symmetric with respect to quarter filling, and the 
unpolarized case (lower curve), whose total energy per site is 
symmetric with respect to half filling.  (see text).}
\label{fig4}
\end{figure}

In Figs.~\ref{fig1}-\ref{fig4} we show the electronic ground-state 
energy per site, obtained by FNMC runs as a function of the electron 
density \cite{apbc}.  Energies are in units of the hopping parameter 
$t$ throughout this paper; the statistical errors are smaller than the 
marker size, and thus are not visible.  The calculated points are 
shown as full markers for closed shells, and as empty markers for open 
shells.  From the comparison of Fig.~\ref{fig1} and Table~\ref{table1} 
it appears that the open-shell error, which we found to be significant 
for a small $4\!\times\!  4$ lattice, becomes of the order of the 
statistical error (and thus negligible \cite{maddeche}) for our large 
lattices \cite{shellstorte}.

\subsubsection{Accuracy}
\label{precisione}

At all densities our three sets of data for $U/t\!=\!10$ 
(Fig.~\ref{fig1}), $20$ (Fig.~\ref{fig2}), and $40$ (Fig.~\ref{fig3}) 
are bracketed, as seen in the cumulative Fig.~\ref{fig4}, by the 
noninteracting unpolarized energy and the fully spin-polarized energy 
(both dashed in Fig.~\ref{fig4}), and display a smooth and reasonable 
behavior.  To evaluate the absolute accuracy of our results, we can 
rely on two exact limits: the low-density ($n\!\simeq\!0$) regime, 
where we expect ${\cal E}=-4n$, and the half filled case ($n\!=\!1$), 
for which the strong-coupling expansion provides the correct 
large-$U/t$ behavior: to leading order in $t/U$, the model maps onto 
an \keyword{Heisenberg model}, whose ground-state energy has been 
evaluated with great accuracy \cite{CASO98,SO93}.  We can also 
consider the next correction term $34.6 t^{4}/U^{3}$ \cite{Taka77}.  
At low density our results are essentially exact; at half filling our 
error is small ($\sim\!  3\%$) for $U/t=10$ but (as already noticed in
Table~\ref{table1}) it tends to grow with $U/t$: for $U/t=20$ it's 
$\sim\!  9\%$, and for $U/t\!=\!40$ it's $\sim\!  11\%$.  We have made 
sure (see markers other than dots in Figs.~\ref{fig1},\ref{fig3} and 
Refs.~\cite{apbc,shellstorte}) that such an energy discrepancy is not 
due to shape, open-shell, finite-size \cite{maddeche}, and boundary 
condition effects; as far as systematic errors are concerned, we are 
thus left with the fixed-node approximation: as $U/t$ grows, more 
flexible trial wavefunctions may be required to obtain more accurate 
energies \cite{long}.

\subsubsection{Phase separation}

\begin{figure}
\begin{center}
\epsfxsize=200 pt
\centerline{\epsfbox{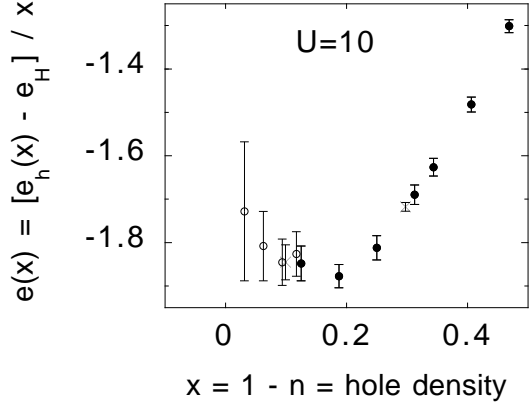}}
\end{center}
\caption[xxx]{Plot of $e(x)$ vs.  $x$ for $U/t=10$.  Full dots correspond to 
closed shells and empty dots correspond to open shells of a $16 
\!\times\!  16$ lattice.  Crosses (corresponding to a $11\sqrt 2\times 
11\sqrt 2$ lattice) are shown for comparison.  Obviously at small $x$ 
the error bar associated to $e(x)$, $\Delta{e(x)} = [\Delta{e}_{h}(x) 
+ \Delta{e}_{h}(x\!=\!0)]/x$, becomes significant even if the 
statistical FNMC error $\Delta{e}_{h}(x)$ is tiny.}
\label{fig5}
\end{figure}

\begin{figure}
\begin{center}
\epsfxsize=200 pt
\centerline{\epsfbox{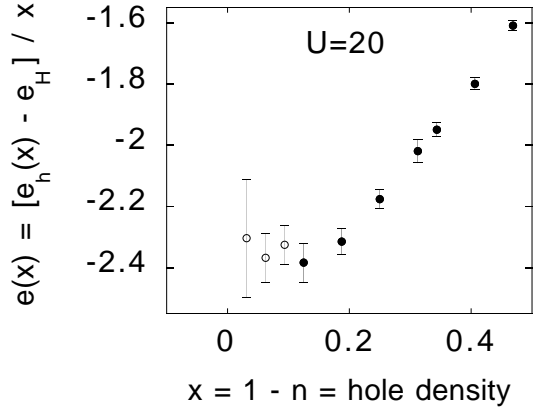}}
\end{center}
\caption{Plot of $e(x)$ vs.  $x$ for $U/t=20$.  Full dots correspond to 
closed shells and empty dots correspond to open shells of a $16 
\!\times\!  16$ lattice.  Other comments like in the previous figure.}
\label{fig6}
\end{figure}

\begin{figure}
\begin{center}
\epsfxsize=200 pt
\centerline{\epsfbox{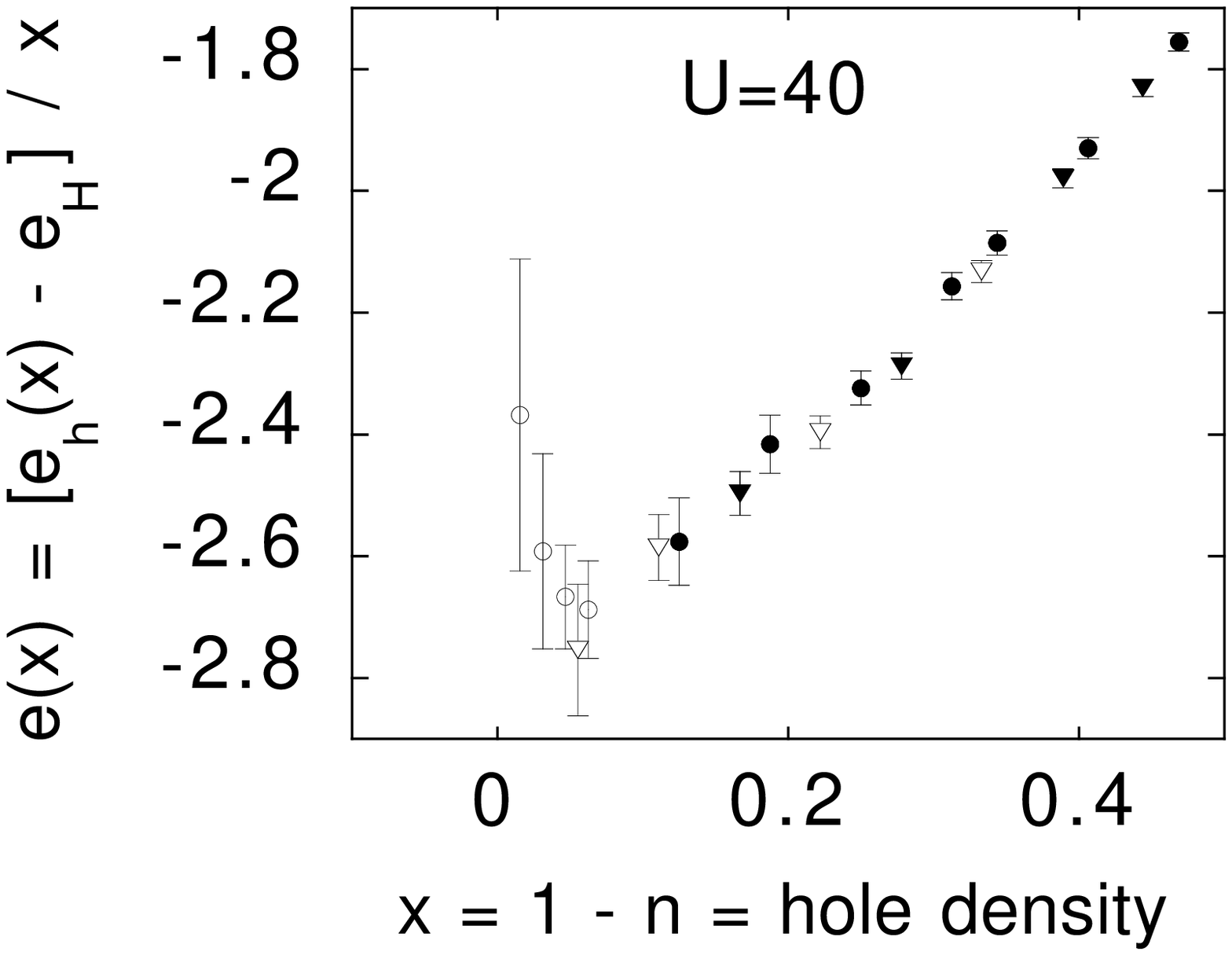}}
\end{center}
\caption{Plot of $e(x)$ vs.  $x$ (see text) for $U/t=40$.  Full dots 
correspond to closed shells and empty dots correspond to open shells 
of a $16 \!\times\!  16$ lattice.  Full and open triangles correspond 
to closed and open shells of a smaller $12 \!\times\!  12$ lattice and 
are shown for comparison (see text).  Other comments like in the 
previous figure.}
\label{fig7}
\end{figure}

Keeping in mind the virtues and limitations of our numerical study, we 
can now turn to the question of phase separation in the Hubbard model.  
In some sense the function $e(x)$ works like a magnifying lens of the 
phase separation (see Fig.~\ref{fig0}).  It should be stressed that in 
a consistent definition of $e(x)$ the half-filling energy $e_{H}$ must 
be obtained as $e_{h}(x \!=\!0)$ from the same calculation as any 
other $e_{h}(x\neq 0)$ (here, from the FNMC at half filling).  If 
that's not the case (for example, if the Heisenberg value is used 
instead), then $e(x)$ will spuriously diverge near $x \!=\!0$, with a 
good chance of artificially creating, rather than magnifying, the 
occurrence of phase separation.  In Figs.~\ref{fig5}-\ref{fig7} we 
find plots of $e(x)$ for $U/t=10$, $20$, and $40$; these values, as 
well as the associated error bars, are directly obtained from those of 
Figs.~\ref{fig1}-\ref{fig3} (i.e.  from the original FNMC energies and 
tiny error bars).  Despite the error bars, a common trend is evident 
for all the calculated coupling strenghts: $e(x)$ has a positive slope 
for large hole densities, far from half-filling, but then it clearly 
changes slope below some small critical density $x_{c}$.  Such a 
minimum in $e(x)$ implies that, at least for the FNMC effective 
hamiltonian determined by our choice of wavefunction and cell size 
\cite{maddeche}, phase separation occurs below $x=x_{c}$.  Although a 
finer grid of hole densities would be required to locate with high 
precision the critical density $x_{c}$ as a function of $U/t$, we 
already see that $x_c$ decreases as $U/t$ is increased; this 
qualitatively agrees with the original predictions \cite{emery2} and 
with some previous calculations on the $t\!-\!J$ model at 
corresponding values of $J = 4t^2/U$ \cite{hellberg}.

\section{Summary, open problems, conclusions and perspectives}
\label{Conclu}

\subsection{Summary}

In summary, the extensive fixed-node Monte Carlo simulations of the 
Hubbard model for $16\!\times\!16$ two-dimensional lattices by 
Cosentini {\it et al.} \cite{cosentini}, just recalled in the previous 
Section, suggest the presence of a phase separation for $U \gg t$.  If 
confirmed, this result would imply that the $t\!-\!J$ model is also 
likely to show a phase separation in the physically relevant regime 
$\!J < \!0.4$, and that even a single-band Hubbard model is sufficient 
to reproduce this physical tendency of some high-$T_c$ superconductors 
(Subs. \ref{motivi}).  Before these statements are confirmed beyond any 
conceivable doubt, further tests may be required because of the 
following two problems.

\subsection{Open problems}
\subsubsection{Fixed-node error at large $U/t$}

First of all, the fixed-node approximation is a variational 
approximation.  Especially for the two larger $U/t$ values, the energy 
discrepancy at half filling (as discussed in Subs.~\ref{precisione}) 
is unfortunately significant.  Not knowing in advance the density 
dependence of such an energy discrepancy, even the conclusions on 
phase separation, based on the function $e(x)$, could be at risk.  
Richer variational wavefunctions (yielding different nodal topologies) 
\cite{long}, or possibly the stochastic reconfiguration scheme 
recently proposed by Sorella \cite{sandro}, must be employed to go 
beyond our simple Gutzwiller-Slater nodes, and to settle this point 
completely.

\subsubsection{Finite-size and ``phase separation'' at $U\!=\!0$}
\label{okkio}

\begin{figure}
\begin{center}
\epsfxsize=200 pt
\centerline{\epsfbox{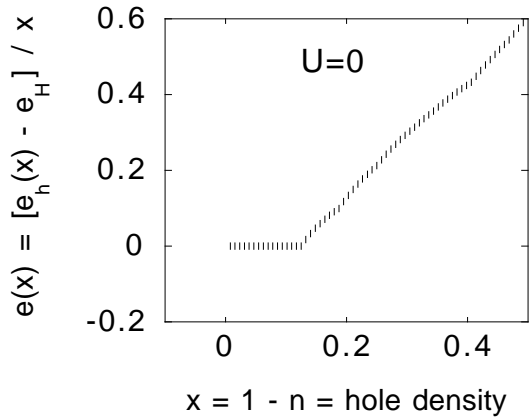}}
\end{center}
\caption{This figure contains plot of $e(x)$ vs.  $x$ for $U/t=0$ 
(noninteracting electrons, analytic results) for a $16 \!\times\!  16$ 
lattice.  As a marker we use here vertical bars; we display data for 
all the 128 densities which can be obtained at this lattice size with 
an even number of electrons. Note that below $x=2/16$ $e(x)$ is 
completely flat.}
\label{fig8}
\end{figure}

\noindent A second motivation for further tests is illustrated in 
Fig.~\ref{fig8}.  Here we show the energy density of a finite 
$16\!\times\!16$ lattice with $U/t=0$ (noninteracting electrons).  
Close to half filling, below $x_{c}=0.125$, the function $e(x)$ 
becomes completely flat.  Such a behavior, already pointed out by Lin 
a few years ago \cite{size} but surprisingly forgotten in most 
subsequent numerical studies of small Hubbard lattices, is an artifact 
related to the high degeneracy of the Fermi level, and disappears in a 
truly infinite lattice ($x_{c}$ vanishes as $2/L$ when $L$, the side 
of a square $L\!\times\!L$ lattice, goes to infinity).  Comparing 
Fig.~\ref{fig8} with previous Figs.~\ref{fig5}-\ref{fig7}, one may 
even fear that the artificial flattening at $U=0$ has something to do 
with the behavior found for $e(x)$ at $U>0$.  This observation 
suggests caution even if, as stated in various parts of 
Subs.~\ref{linummeri}, many severe tests of open-shell and finite-size 
effects have already been passed.  The almost perfect convergence of 
energies ${\cal E}(n)$, shown in Figs.~\ref{fig1}-\ref{fig3}, and even 
the convergence (within the appropriate statistical error) of the 
$e(x)$ data of Figs.~\ref{fig5}-\ref{fig7}, could, incredibly, not be 
enough to rule out a very slow $2/L$ drift of the critical density 
towards zero.  Perturbation theory or further numerical work at small 
$U/t$, as well as a thorough finite-size scaling study based on yet 
larger lattices may be required to convincingly rule out the 
possibility of such a spurious effect.
 
\subsection{Conclusions and perspectives}

Neither previous, quite crude numerical studies, nor the recent (and 
definitely much more accurate) work by Cosentini {\it et al.}, 
recalled in this chapter, seem to be sufficient for a 100\%
safe and general claim on phase separation.  This is by itself 
information of the greatest importance to whoever is active either in 
the theory or in the numerical simulation of strongly correlated 
two-dimensional electron systems: phase separation is a very delicate 
effect, which can be masked by otherwise negligible variational or 
finite-size errors.  The simplicity and ease of the lattice fixed-node 
Monte Carlo \cite{bemmel,ceperley}, the efficiency of the new 
Calandra-Sorella reconfiguration scheme \cite{CASO98}, and the 
experience gained in the preliminary study recalled here 
\cite{cosentini} provide a good starting point and very useful 
guidelines for a future, conclusive numerical study of this 
challenging problem.

\section{Acknowledgements}
Thanks are due to M. Capone and L. Guidoni for many stimulating 
discussions during the preparation of both the lectures and the 
manuscript, and to M. Calandra Buonaura, C. Lavalle and G. Senatore 
for patiently listening to GBB's rehearsals in Ithaca, and for their 
useful comments .  Partial support from the Italian National Research 
Council (CNR, Comitato Scienza e Tecnologia dell'Informazione, grants 
no.  96.02045.CT12 and 97.05081.CT12), the Italian Ministry for 
University, Research and Technology (MURST grant no.  9702265437) and 
INFM Commissione Calcolo is gratefully acknowledged.


\begin{thebibliography}{}
\bibitem{realism} In this respect it may be worth mentioning that, 
although model hamiltonians can be intrinsically exciting to some 
theorists \cite{actually}, their link to real bands and bonds, 
following a pioneering paper by Andersen and Jepsen \cite{andjeps}, 
can in fact be reasonably worked out, at least as far as the 
one-electron matrix elements are concerned \cite{koch_book,andersen}.
\bibitem{actually} The work presented in the second part of this 
chapter belongs in some sense to this line of thought: it's based on 
the simplest 2D Hubbard hamiltonian (just one orbital per site) and 
unrealistically large values of the electron-electron repulsion.  
However, as discussed in Sect.~\ref{2DHub}, the state of the 
controversy on phase separation is such that even an oversimplified 
model like this might give a significant contribution, and possibly 
shed some light even on real systems.
\bibitem{digression} A common option is ${\cal G}=1-\tau\left({\cal
H}-E_{ref}\right)$, with a convenient choice of a small positive
$\tau$ and of the energy reference $E_{ref}$.  This operator looks
like a short-time Green's function, but no short-time approximation is
involved here: on a lattice this is just a way of constructing an
iterative ground-state projector, because the spectrum of the
hamiltonian is bounded both from above and from below.
\bibitem{excited} Under the conditions of the previous 
Ref.~\cite{digression}, analogous projectors can be constructed for 
the highest excited state by e.g.  reversing the sign of $\cal H$; in 
this chapter, however, we only deal with ground-state projectors.
\bibitem{sparse} For example, in the Hubbard model of this chapter, the 
number of basis states $\ket{X}$ connected to a given state $\ket{X'}$ 
- and thus the number of operations needed to evaluate, say, 
$\bra{X}{\cal H}\ket{\Psi}$ - grows linearly with the number of 
electrons $N_{e}$; which in turn, because of the Pauli principle, can be 
at most twice the number of sites $N_{s}$.  This is fortunately much less 
than the size of the Hilbert space, which, for $N_{e}$ identical particles 
on a lattice of $N_{s}$ sites and an unspecified total spin $S$ and 
projection $S_{z}$, is $(2N_{s})!/(2N_{s} \!-\!  N_{e})!$ for ``labeled 
fermion'' configurations (see e.g.  the appendix B of 
Ref.~\cite{ceperley}), and $2N_{s} \choose N_{e}$ for (antisymmetric) true 
fermion states.  In both cases, the overwhelming majority of 
hamiltonian matrix elements are zero (sparse matrix).
\bibitem{averages} For example, one could estimate the largest 
eigenvalue of $\cal G$ as $\gamma$ = $\sum_i w_i(L) b_i / \sum_i 
w_i(L)$; then, if the relation between $\cal G$ and $\cal H$ follows 
the choice of Ref.~\cite{digression}, the estimate for the 
ground-state energy would be $E_{\rm o}=\left[ 
\left(1\!-\!\gamma\right) / \tau \right] + E_{ref}$.
\bibitem{secondqua} The following digression could be useful to some 
reader.  The Hubbard hamiltonian is usually written in terms of 
creation and annihilation operators, according to the formalism of 
second quantization (see e.g.  Eq.~\ref{hubbardham}); in this context 
each of the basis states $\ket{X}$ of the many-body Hilbert space is 
identified by an array of occupation numbers, which in turn 
corresponds to a Slater determinant of occupied single-particle 
spin-orbitals.  Each individual basis state $\ket{X}$, and thus any 
linear combination of these basis states, is therefore an 
automatically antisymmetric fermion function.  With this choice the 
``sign problem'' is built into the basis set, and directly emerges, 
without the need of any fermionic trial function, from the 
off-diagonal matrix elements: because of the fermionic commutation 
rules of the creation and annihilation operators, some $\bra{X'} {\cal 
H} \ket{X}$ turn out to be equal to $-t$, while others are equal to 
$+t$.  If we instead choose a different (and larger \cite{sparse}) 
basis $\ket{X}$ of ``labeled fermion'' configurations, as explained 
e.g.  in appendix B of Ref.~\cite{ceperley}, then the sign problem 
separates into ``good'' off-diagonal hopping terms (all $\bra{X'} 
{\cal H} \ket{X}$ are now negative and equal to $-t$) and a ``bad'' 
antisymmetry requirement, which is imposed through a fermionic trial 
function $\Psi_{T}(X)$, and causes sign flips in the 
importance-sampled Green's function matrix $ \tilde{\cal G}_{X'X} = 
\Psi_{T}(X') \bra{X'} {\cal G} \ket{X}\Psi_{T}^{-1}(X)$.  The latter 
choice corresponds to an intuitive picture which parallels the 
continuum case and may be convenient for practical purposes, but it's 
inconsistent with the usual second quantization.
\bibitem{lever} It also gives a sound proof of the variational 
principle and withdraws an incorrect statement of Ref.~\cite{bemmel} 
about ``lever rules''.
\bibitem{singlet} This choice corresponds to a projection of the total 
spin $S_{z}=0$ and, for those electron densities which correspond to a 
closed-shell situation, also to a total spin $S=0$.
\bibitem{apbc} Different boundary conditions yield energies within an
error bar for $16 \!\times\! 16$ lattices, and within at most two
error bars for our smallest lattices; all results presented here refer
to antiperiodic boundary conditions.
\bibitem{maddeche} We will come back to this point while discussing 
our results in Subs.~\ref{okkio}.
\bibitem{shellstorte} We have checked a few densities which correspond 
to open shells in our usual $16 \!\times\!  16$ lattice (empty markers 
in Figs.~\ref{fig1}-\ref{fig7}), but to closed shells in a 
$45^{o}$-rotated $11\sqrt 2 \!\times\!  11\sqrt 2$ lattice (crosses in 
Figs.~\ref{fig1},\ref{fig4} and \ref{fig5}); the energy difference 
turns out to be negligible \cite{maddeche}.
\bibitem{long} The Gutzwiller-Slater wavefunction, a very popular 
choice for the Hubbard model, suffers from at least two drawbacks 
(which probably reflect into the ``quality'' of its nodes): for open 
shells it does not describe a spin singlet, and it only has on-site 
correlations; more general forms, such as those proposed in 
Ref.~\cite{lhuillier}, could gain some of the missing energy at 
$U/t=20,40$.
\end{thebibliography}
\end{document}